# Cosmic Ray Spectral Deformation Caused by Energy Determination Errors

Per Carlson[*1] and Conny Wannemark[*2]
*Physics Department, Royal Institute of Technology (KTH), 10691 Stockholm, Sweden*

## Abstract

Using simulation methods, distortion effects on energy spectra caused by errors in the energy determination have been investigated. For cosmic ray proton spectra, falling steeply with kinetic energy E as $E^{-2.7}$, significant effects appear. When magnetic spectrometers are used to determine the energy, the relative error increases linearly with the energy and distortions with a sinusoidal form appear starting at an energy that depends significantly on the error distribution but at an energy lower than that corresponding to the Maximum Detectable Rigidity of the spectrometer. The effect should be taken into consideration when comparing data from different experiments, often having different error distributions.

## 1. Introduction

A good knowledge of the steeply falling cosmic ray spectra of protons and nuclei is essential for e.g. calculations of the interstellar production of antiprotons, positrons and photons. The calculation of the atmospheric production of neutrinos is also crucially dependent on a good knowledge of the flux of cosmic ray protons and nuclei. The determination of the energy of individual cosmic ray particles is subject to random statistical errors as well as to systematic errors. When the statistical error is no longer small compared to the energy bin used, the steeply falling spectrum may cause deformations in the measured one. This feature is also true for spectra of photons and other charged particles.

It was early observed [1-5] that the steep form of the cosmic ray proton spectrum, where the particle flux decreases with kinetic energy E approximately as $E^{-2.7}$, together with a non-negligible error in the determination of the energy, causes a systematic overestimate of the energy.

There are different ways of determining the particle energy in cosmic ray experiments. A magnetic spectrometer gives a relative momentum error $\Delta p/p$ for charged particles that is proportional to p. In some experiments the amount of Cherenkov light is used to determine the energy. In this case the relative energy $\Delta p/p$ increases as $p^2$, assuming that the error is determined from the number of observed Cherenkov photons. In air shower experiments, used for very high energies, the energy dependence of the error seems weaker than with other methods, and a logarithmic dependence has been used [5].

It is clear that when the energy resolution is much smaller than the bin width used to present data, the effect of the energy resolution can be neglected. However, many experiments tend to use as much as possible of accumulated data, including events with even 50 % or worse energy resolution. In this case there are significant effects on the observed spectrum from the resolution.

In this paper we present results of a simulation study where effects of the energy resolution is studied for typically binned data, focussing on experiments using magnetic spectrometers and for energies up to about 1 TeV. Section 2 presents earlier work, section 3 lays out the simulation procedure and the results are presented in section 4. A summary is given in section 5.

**2. Earlier work**
A detailed analysis of the effect of errors on high energy air shower data was done by Edge et al. [5]. They conclude, from a simulation study assuming an error in the energy determination of 13 % at $10^{17}$ eV increasing logarithmically to 200 % at $10^{19}$ eV, that the spectrum at lower energies correctly describes the true spectrum whereas at energies closer to $10^{19}$ eV the spectrum flattens significantly.

A method to correct measured energy spectra of cosmic ray nuclei in the range up to about 100 GeV/nucleon is discussed by Juliusson [6]. The measurements used the Cherenkov technique and the energy was estimated from the number of observed photons. The relative error in the momentum $\Delta p/p$ is in this case proportional to $p^2$. Juliusson concludes that measured fluxes must be increased by about 10 % to as much as a factor of 2 for energy bins where the measured momentum is greater than about twice the Cherenkov threshold. The method was recently used [7] to correct measured cosmic ray fluxes of protons and Helium nuclei in the range 30 – 150 GeV/nucleon.

Spectrum deformation effects caused by the energy resolution has been discussed for the BESS experiment [8] that utilizes a cylindrical magnetic spectrometer. In the BESS-98 experiment the deflection uncertainty distribution peaks at 5 $(TV)^{-1}$ corresponding to a Maximum Detectable Rigidity MDR (Rigidity R = pc/eZ, where p is the momentum, c the velocity of light and eZ the particle charge) of 200 GV [9]. From simulation studies it is concluded that the proton spectrum deformation is small below about 120 GeV whereas at 230 GeV the measured spectrum is 20% larger than the input $E^{-2.8}$ proton spectrum. For the improved BESS-TeV spectrometer [8], where the MDR is much higher, 1.3 TV, the deformation [10] for protons and helium nuclei (muons) was studied using an $E^{-2.7}$ ($E^{-3.2}$) input spectrum.

The deformation was found to be less than 5 % below 1 TV (400 GV). The larger effect for muons is caused by the larger spectrum index of – 3.2 as compared to – 2.7 for protons and helium nuclei.

Because of steeply falling fluxes, data are often presented as fluxes multiplied with $E^{2.75}$ or $E^{2.5}$ which makes the plots easier to show. When comparing data from different experiments care must be taken to estimate the effects of the error of the energy determination, since errors assigned to each data point in these plots are usually based on the number of observed events, neglecting the errors coming from the energy. Note that a 5% scale error in the energy determination changes the flux multiplied by $E^{2.75}$ by 14%. Detector inefficiencies, on the other hand, change fluxes linearly.

Finally we mention that the problem of where to stick the data points in wide bins is discussed in a paper by Lafferty et al. [11].

**3. The simulation procedure**
Many experiments, most of them using magnetic spectrometers, have published results of cosmic ray proton flux measurements around 100 GeV (for a summary of available data see, e.g., Boezio et al. [12] or Haino et al. [10]). Results from different experiments given as flux×$E^{-2.75}$ differ by about 20%. With the aim to examine in what respect errors of the energy determination might influence the measured spectrum, we choose as an input a cosmic ray proton spectrum of the form dN/dE = const × $E^{-2.75}$ (with E the kinetic energy). We also limited the study in this work to simulate errors in magnetic spectrometers, i.e. $\Delta E/E$ = const.×E (for kinetic energies much larger than the proton mass).

In many cosmic ray magnetic spectrometers the magnetic field is not homogeneous with the BESS spectrometer being an exception. The inhomogeneity introduces an asymmetric distribution of deflection uncertainty with a tail for large deflections.

As a first step we compared simulations using a deflection error from the CAPRICE98 experiment [12] folded with a normal (Gaussian) error distribution with simulations using instead the average of the CAPRICE98 deflection distribution folded with the normal error distribution. There were no significant changes in the resulting spectra. Here we therefore use the average of the CAPRICE98 deflection uncertainty distribution giving a relative error $\Delta E/E$ = 0.005×E (E in GeV), corresponding to a MDR of 200 GV. The average deflection uncertainty was folded with a normal or log-normal error distribution, the latter giving an asymmetric distribution characteristic of random errors changing the result with a multiplicative factor and always giving positive energies. We also present results using a smaller relative error, $\Delta E/E$ = 0.001×E, corresponding to a MDR of 1 TV.

A proton was selected from the CAPRICE98 flux spectrum [12]. Its kinetic energy E was given an error following the normal or log-normal distribution. The simulated protons were grouped in energy bins. Finally the resulting flux in each bin was multiplied by $E^{2.7}$ where E was chosen according to the method in Ref. [11].

**4. Results and discussion**

The results of the simulations are shown in Fig. 1-4. In Figs. 1 and 2 the errors are normally distributed and results of two values of the relative error are given, $\Delta E/E = 0.005 \times E$ (approximately the CAPRICE98 average deflection uncertainty) and $\Delta E/E = 0.001 \times E$ (E in GeV). Fig. 1 shows the flux multiplied with $E^{2.7}$ and Fig. 2 the relative deformation. Fig. 3 and 4 show the corresponding results using instead log-normal distribution for the errors. In Fig. 1 and 3 we also show the CAPRICE98 results.

For $\Delta E/E = 0.005 \times E$ the simulations follow the input spectrum until about 30 (50) GeV for the normal (log-normal) distribution, corresponding to $\Delta E/E$ = 15% (25%). For higher energies there is a sinusoidal form with a minimum about 15% below the input at 120 (200) GeV followed by a crossing of the input at 350 (1100) GeV, the latter effect show up because the flux is multiplied by $E^{2.7}$ For even higher energies the simulations give fluxes larger than the input. We note that the peak of the CAPRICE98 deflection distribution corresponds to a MDR of 350 GV [12]. If instead the average of the deflection distribution is used for CAPRICE98, because of the long tail for large deflections, the corresponding MDR is 200 GV.

The decrease in simulated flux above 30 (50) GeV followed by an increase above 350 (1100) GeV in Fig. 1 is explained as follows. The number of events in an energy bin is the number of simulated events in that bin plus influx and minus outflux. Since the error in energy increases as the energy squared, the influx and outflux will be different for different energy bins. The error increases as the energy squared and with increasing energy, at some point the negative error in a given energy bin becomes so large that the outflux from that bin will fall not into the next lower energy bin but into the next-to-next lower energy bin causing the decrease. The increase above 350 (1100) GeV is caused by an influx from lower energy bins that is larger than the outflux. Because the log-normal distribution is asymmetric with short tails for negative errors, the distortion effects of the spectrum will be moved to higher energies.

The energy where the distortions start depends crucially on the energy distribution. Our conclusion is that there are significant effects on the simulated spectrum. These effects depend critically on the error distribution, being significant for normally distributed errors even below 50 GeV. For log-normally distributed errors the effect is about 10 % at 100 GeV. One should bear in mind that in real experiments the true error distribution can be more complicated and our results should therefore be taken with care.

For $\Delta E/E = 0.001 \times E$ the simulations follow the input until 120 (200) GeV for normally (log-normally) distributed errors, corresponding to $\Delta E/E = 10\%$ (20%). There is a minimum at 600 (1000) GeV followed by crossing the input spectrum at 2 (6) TeV. We note that the BESS-TeV (MDR 1.4 TV) spectrum deformation study [10] shows that the input spectrum is followed until about 300 GeV.

For clarity we have chosen to include in the figures only the experimental data from the CAPRICE98 experiment. The recent data from the BESS-TeV spectrometer [10] with a MDR of 1.4 TV covers the energy range up to about 500 GeV that corresponds to 36% of the MDR. The CAPRICE98 data covers the range up to 300 GeV, corresponding to 86% of the MDR calculated from the peak of the deflection distribution. The CAPRICE98 data seem to decrease faster with energy than the BESS data above 100 GeV. The deformation effect shown in our simulations could explain this effect. Fitting the CAPRICE flux to the form const × $E^{-\alpha}$ gives $\alpha = 2.75 \pm 0.02$ for energies above 20 GeV but $\alpha = 2.80 \pm 0.02$ for the high energy range 40 – 350 GeV giving support for this conjecture.

Our simulations have been limited to magnetic spectrometers. However, deformation effects can appear also in other types of spectrometers and care must be taken in interpreting results. We note, e.g., that in a recent study [13] of the gamma-ray spectrum from the centre of our galaxy, data from the EGRET experiment with energies over 100 GeV have been included, although the error on the energy estimate is about 50% or larger. Deformation effects, demonstrated in this paper for magnetic spectrometers, could also be important for the photon data and alter conclusions based on the shape of the spectrum.

## 5. Summary
We show from simulations that systematic spectral deformation effects appear when steeply falling cosmic ray spectra are measured with magnetic spectrometers where the relative error is proportional to the energy: $\Delta E/E = a \times E$. The deformation of sinusoidal form starts at an energy that depends strongly on the form of the error distribution and on the value of a. For normally distributed errors with a = 0.005, corresponding to a MDR of 200 GV, the deformation starts at 30 GeV, whereas for log normally distributed errors the start is at 50 GeV. After a decrease of about 15% the simulated spectra increases above the input spectrum.

We note that many measurements of cosmic ray energy spectra using magnetic spectrometers extend well into energies where deformation effects become significant. Deformation effects also appear using other types of energy determination as e.g., air shower arrays and space borne calorimeters.

**Acknowledgements**
We thank the Swedish Space Board for support.

[*1]carlson@particle.kth.se
[*2]connyw@kth.se

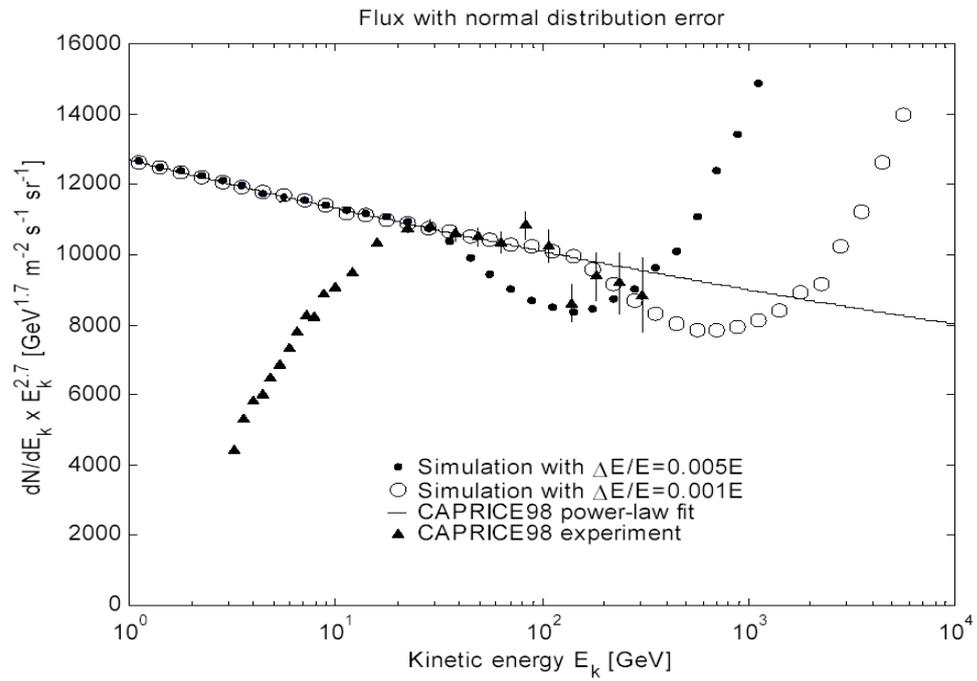

Fig. 1. The proton flux multiplied with $E^{2.7}$ as function of the kinetic energy. Data on protons from the CAPRICE98 experiment are shown together with the power-law fit from that experiment. Simulation results are shown using normally distributed errors with $\Delta E/E = 0.005 \times E$ and $\Delta E/E = 0.001 \times E$ with E in GeV.

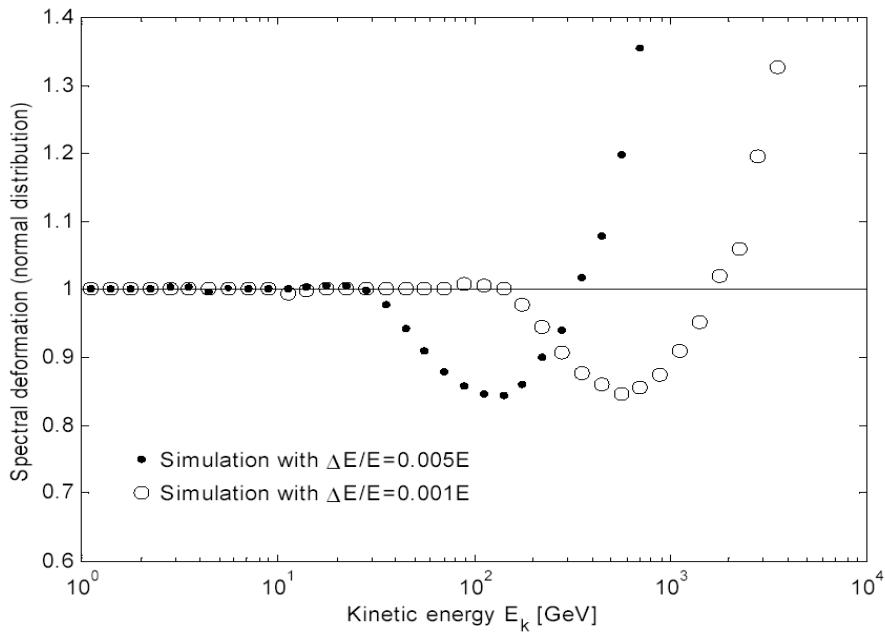

Fig. 2. Spectral deformation defined as the ratio between input and simulated spectrum for ΔE/E = 0.005×E and ΔE/E = 0.001×E with E in GeV. Normally distributed errors.

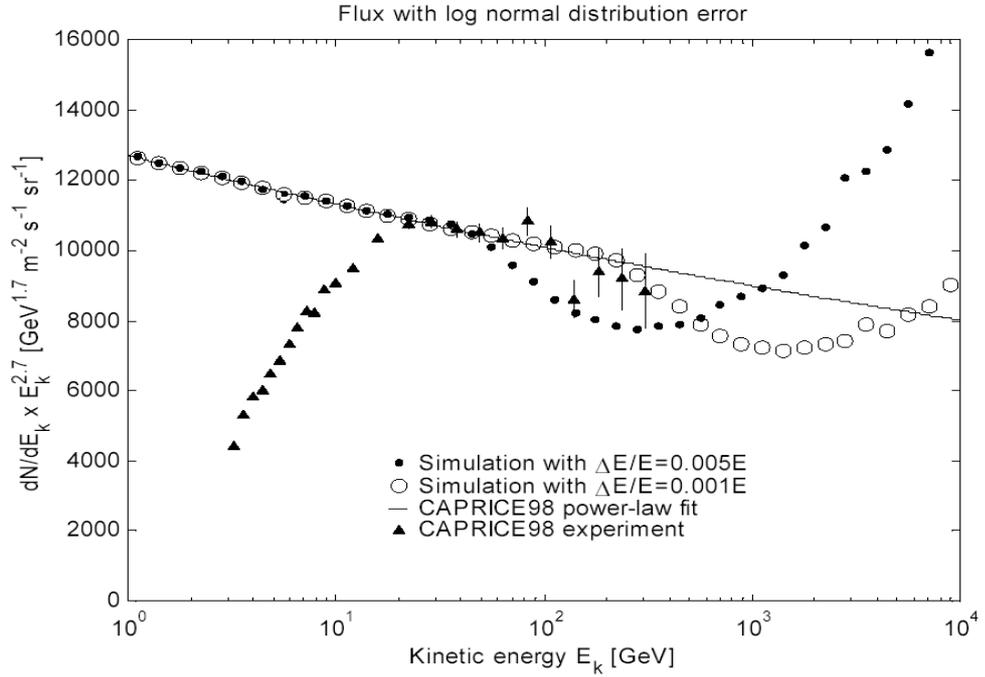

Fig. 3. As for Fig. 1 but with errors that follow a log-normal distribution and ΔE/E = 0.005×E and ΔE/E = 0.001×E with E in GeV.

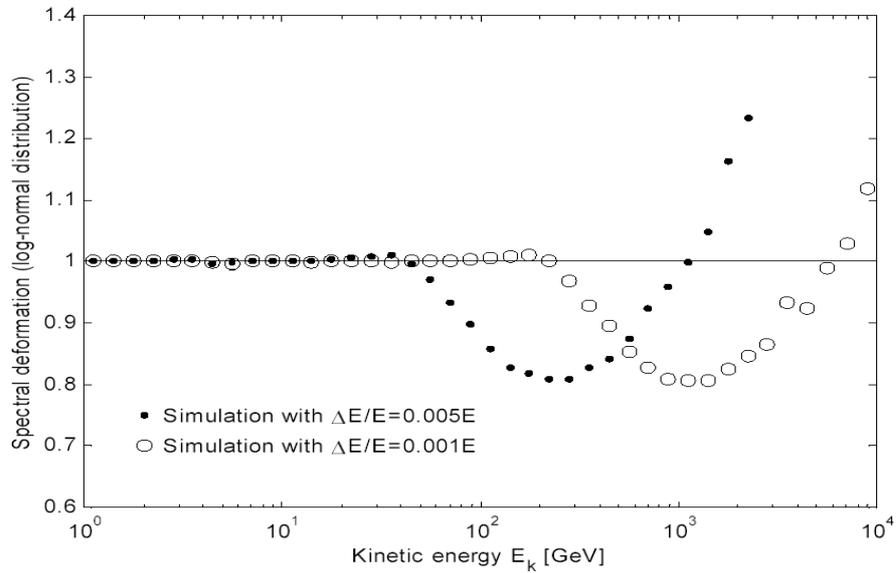

Fig 4. Spectral deformation defined as the ratio between input and simulated spectrum for ΔE/E = 0.005×E and ΔE/E = 0.001×E with E in GeV. Log-normally distributed errors.